%% file: paper.tex
\documentclass[preprint,journal]{vgtc}       

\ifpdf
  \pdfoutput=1\relax                   
  \usepackage{graphicx}                
  \DeclareGraphicsExtensions{.pdf,.png,.jpg,.jpeg} 
\else
  \usepackage{graphicx}                
  \DeclareGraphicsExtensions{.eps}     
\fi%

\graphicspath{{figures/}{pictures/}{images/}{./}} 

\usepackage{amsmath}
\usepackage{amssymb}
\usepackage{microtype}
\usepackage[super]{nth}
\usepackage[section]{placeins}
\usepackage{subfig}
\PassOptionsToPackage{warn}{textcomp}  
\usepackage{textcomp}                  
\usepackage{mathptmx}                  
\usepackage{times}                     
\usepackage{cite}                      
\usepackage{tabu}                      
\usepackage{booktabs}                  
\usepackage{wrapfig}
\newcommand{\eat}[1]{}
\newcommand{\mb}[1]{{\mathbf{#1}}}

\preprinttext{To appear in IEEE Transactions on Visualization and Computer Graphics.}

\onlineid{1178}

\vgtccategory{Research}
\vgtcpapertype{Application/Design Study}

\title{Attention Flows: Analyzing and Comparing Attention Mechanisms in Language Models}

\author{Joseph F DeRose, Jiayao Wang, 
and Matthew Berger}
\authorfooter{
\item Authors are with Vanderbilt University. Corresponding e-mails: \{joseph.f.derose, jiayao.wang, matthew.berger\}@vanderbilt.edu.
}

\shortauthortitle{DeRose \MakeLowercase{\textit{et al.}}: Attention Flows: Analyzing and Comparing Attention Mechanisms in Language Models}

\abstract{Advances in language modeling have led to the development of deep attention-based models that are performant across a wide variety of natural language processing (NLP) problems. These language models are typified by a pre-training process on large unlabeled text corpora and subsequently fine-tuned for specific tasks. Although considerable work has been devoted to understanding the attention mechanisms of pre-trained models, it is less understood how a model's attention mechanisms change when trained for a target NLP task. In this paper, we propose a visual analytics approach to understanding fine-tuning in attention-based language models. Our visualization, Attention Flows, is designed to support users in querying, tracing, and comparing attention within layers, across layers, and amongst attention heads in Transformer-based language models. To help users gain insight on how a classification decision is made, our design is centered on depicting classification-based attention at the deepest layer and how attention from prior layers flows throughout words in the input. Attention Flows supports the analysis of a single model, as well as the visual comparison between pre-trained and fine-tuned models via their similarities and differences. We use Attention Flows to study attention mechanisms in various sentence understanding tasks and highlight how attention evolves to address the nuances of solving these tasks.
} 


\CCScatlist{ 
 \CCScat{K.6.1}{Management of Computing and Information Systems}%
{Project and People Management}{Life Cycle};
 \CCScat{K.7.m}{The Computing Profession}{Miscellaneous}{Ethics}
}

\teaser{
 \centering
 \hspace*{-1cm}
 \includegraphics[width=1.1\linewidth, height=.6\linewidth]{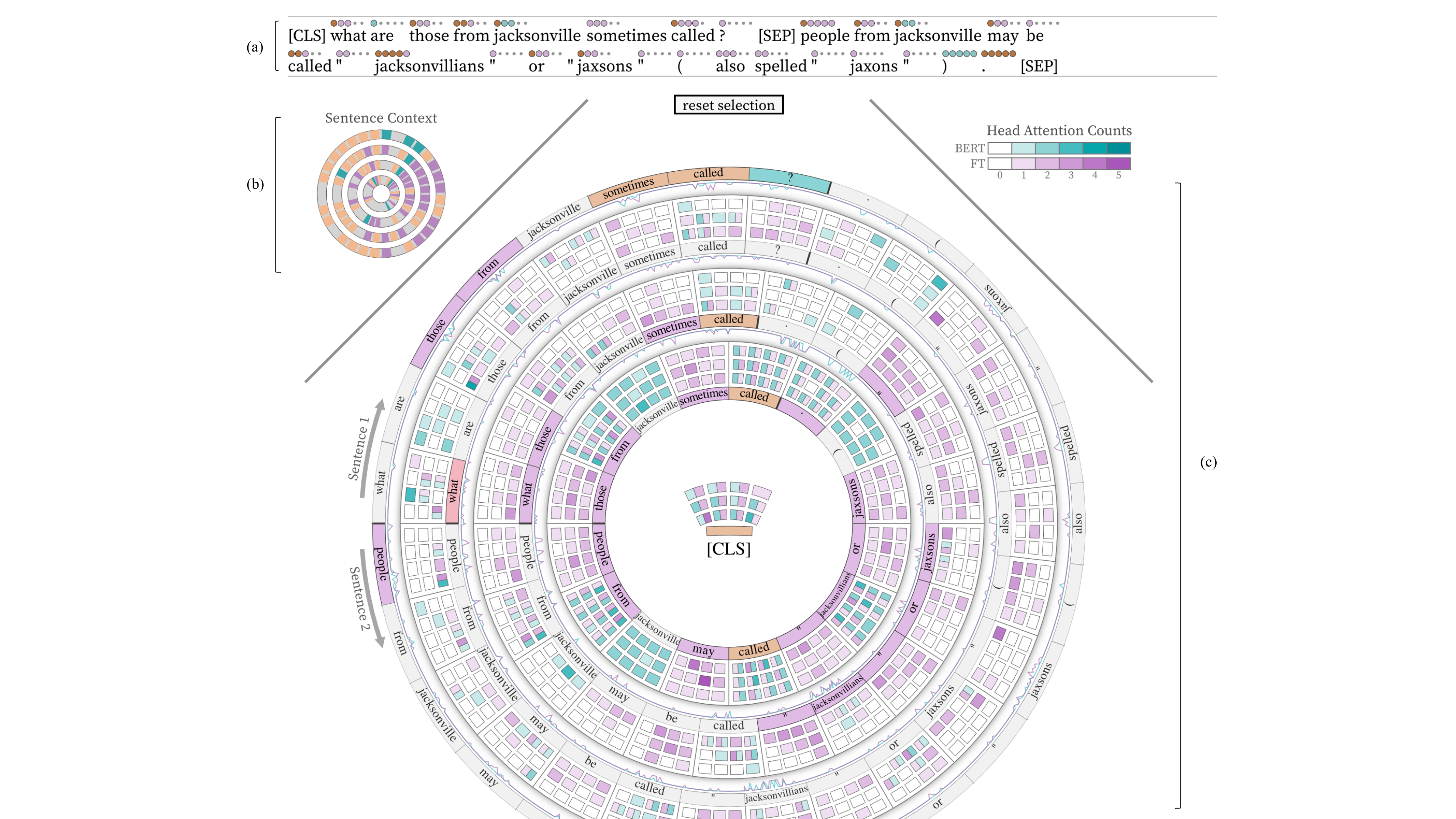}
 \caption{Our approach supports the comparison of attention mechanisms in language models. We compare the BERT model (turquoise) and its fine-tuned counterpart (purple) tasked with determining question-answer pair validity (a). By selecting the word ``what", in contrast to BERT the fine-tuned model attends to the answer ``jacksonvillians or jaxons" (c), with full sentence context shown in (b).}
 \label{fig:teaser}
}

\vgtcinsertpkg

\begin{document}


\firstsection{Introduction}

\maketitle

\input{intro.tex}

\section{Related Work}

\input{related.tex}

\section{Overview and Tasks}

\input{tasks.tex}

\section{Visualization Design}

\input{design.tex}

\section{Experiments}

\input{experiments.tex}

\section{Conclusions and Future Work}

\input{futurework.tex}

\maketitle

\bibliographystyle{abbrv-doi}

\bibliography{refs}
\end{document}

%% file: intro.tex
Recent breakthroughs in natural language processing (NLP) have led to the development of models that yield significant performance gains across a wide variety of language understanding tasks. In particular, BERT~\cite{devlin2019bert} -- or Bidirectional Encoder Representations from Transformers -- has demonstrated how Transformer models~\cite{vaswani2017attention}, pre-trained on unsupervised tasks from large-scale document corpora (e.g. Wikipedia), can be effectively fine-tuned for downstream supervised tasks. Remarkably, the pre-training tasks used in BERT, such as masked word prediction and next sentence prediction, at a glance appear quite different from the downstream tasks, such as question answering, textual entailment, and semantic equivalence of sentences~\cite{wang2018glue}. Although pre-training tasks encourage representations that capture lexical and syntactic reasoning~\cite{clark2019does}, how these language models generalize to sentence understanding tasks remains unclear. This problem is important to model builders, as the relationship between pre-trained and fine-tuned models can help them design pre-training tasks. For instance, identifying differences between models and whether these differences are meaningful to the downstream task at hand, can help ensure that pre-training does not merely learn semantically-irrelevant data nuances that happen to be discriminative for the task~\cite{mccoy2019right}. We see this problem as timely, as recent work has demonstrated the increasing importance of pre-training, resulting in numerous types of unsupervised and weakly-supervised tasks aimed at predicting text spans~\cite{joshi2020spanbert}, entities~\cite{zhang2019ernie}, and word senses~\cite{levine2019sensebert}.

In this work, we seek to obtain a better understanding of BERT models and in particular, to understand the gap between pre-trained models and fine-tuned models. The problem of understanding BERT poses numerous challenges due to its inherent complexity. Specifically, sentence-contextualized word embeddings are learned across multiple layers, and each layer performs so-called \emph{self-attention}, expressing a single word's output embedding as a convex combination of all input word embeddings. Further, self-attention is distributed amongst multiple \emph{attention heads} that operate independently and collectively form a word's representation at a given layer. Recent work has addressed interpretability of self-attention mechanisms, demonstrated only on pre-trained models, showing their ability to capture dependency syntax~\cite{clark2019does} and grammatical relationships~\cite{lina2019open}. Yet these works are typically focused on analyzing a particular relationship known a \emph{priori} and/or studying a single attention head or attention distributed over a given layer. In the context of fine-tuning, our object of study is the classification decision, e.g. does one sentence entail another? Answering this type of question requires a more holistic view of the model, and understanding how attention heads, across multiple layers, aggregate to form a single classification output.

We propose \emph{Attention Flows}: a visual analytics approach to help interpret how classification outputs are formed in BERT models via the visual analysis of attention propagation. Our visualization design is classification-centric: in BERT, classification is performed with respect to a reserved classification token's embedding at the final layer of the model. Thus, the main objective of our visualization is to provide insight on how attention flows across words, both between sequences of layers and within attention heads, down to the eventual classification. We iteratively extract word dependencies between layers using self-attention, working backwards. We extract words at the last layer that most influence classification, and then for each of these words, we extract their respective dependencies from the previous layer, repeating this process throughout all network layers. We visually encode this information in a radial layout: each ring of the layout represents a layer, with the classification token output being at the center and shallower layers progressing outward, shown in Fig.~\ref{fig:teaser}. We encode both words \emph{and} attention heads, where through a set of supported interactions, users can explore how attention flows into, and out of, words.

Critically, our visualization design naturally lends itself towards \emph{model comparison}. We permit the visual comparison of attention flows between a provided pre-trained model and a fine-tuned model, so that the user can comprehend differences and similarities between the models. We show, through use cases and user feedback on a number of sentence understanding tasks~\cite{wang2018glue}, how our visualization highlights distinguishing factors between attention flows of pre-trained models and their fine-tuned counterparts. For instance, we find that question answer inference tasks lead to attention flows that target the ``Five Ws'' for information gathering, while for sentence paraphrasing tasks, we find that when one sentence does not paraphrase another, attention is focused around phrases that differentiate the sentences.


We summarize our contributions below:
\begin{enumerate}
	\item We introduce a visualization design that supports a comprehensive understanding of self-attention in Transformer models, over multiple layers and attention heads. Our design enables inspection of how classification decisions are made in sentence understanding tasks.
	\item Our visualization supports the comparison of self-attention mechanisms between two models, focused on the differences and similarities amongst pre-trained and fine-tuned models.
	\item Through use cases and user feedback, we show how our interface offers insight on changes in self-attention that are due to fine-tuning for inference tasks such as textual entailment, question answering, and sentence paraphrasing.
\end{enumerate}

%% file: related.tex
Our work is related to the visual analysis and interpretability of deep learning, with an emphasis on NLP models. Here we discuss work related to deep networks for NLP, approaches to interpretability, visual analytics for interpreting deep learning, and comparative visualization.

\subsection{Language Modeling}
\label{subsec:ref_lm}

Research in language modeling is focused on learning representations of linguistic elements, typically words and sentences, that capture syntactic and semantic properties suitable for higher-level language understanding. Neural language models, in particular, are focused on learning word~\cite{bengio2003neural} and/or character~\cite{kim2016character} level representations, where the task is to predict a probability distribution over words at a particular point in a sequence, conditioned on all prior words. It is common to use recurrent neural networks (RNNs)~\cite{mikolov2010recurrent}, and variants such as Long Short Term Memory (LSTM)~\cite{hochreiter1997long} or Gated Recurrent Units (GRUs)~\cite{cho2014learning}, for this task to learn contextualized word-level embeddings, given context-independent embeddings as input, e.g. GLOVE~\cite{pennington2014glove} or word2vec~\cite{mikolov2013distributed}. A key aspect of language models is that they do not require human-annotated data, but instead, they learn from document corpora where sentence-level information (e.g. sequences of words) is preserved. Language modeling is usually seen as a \emph{pre-training} process, where contextualized representations are used for downstream supervised NLP tasks, e.g. textual entailment~\cite{bowman2015large}, semantic role labeling~\cite{he2017deep}, and named entity extraction~\cite{peters2017semi}.

An alternative to recurrent models~\cite{peters2018deep} are Transformers~\cite{vaswani2017attention} for modeling sequential data. Transformers rely on a notion of \emph{self-attention}: given input word embeddings combined with positional word encodings, the Transformer outputs a new representation of each word, in part, through a convex combination over all inputs. This convex combination -- represented as a set of nonnegative weights over words that sum to 1 -- assigns importance to words, namely the output embedding of a particular word is dependent on another word's input embedding if its attention weight is high. Further, it is common to employ multiple, independent forms of attention at a given layer through so-called \emph{attention heads}. The approach of BERT~\cite{devlin2019bert} has demonstrated how to use Transformers as language models, via solving the pre-training tasks of masked word prediction, as well as next sentence prediction. They demonstrated that, by \emph{fine-tuning} such pre-trained Transformer models on supervised NLP tasks, strong improvements in performance over a variety of models can be obtained. This has motivated recent work in refining optimization procedures~\cite{liu2019roberta} for Transformers, as well as designing different pre-training tasks, e.g. predicting text spans~\cite{joshi2020spanbert}, entities~\cite{zhang2019ernie}, and word senses~\cite{levine2019sensebert}. The predominance of Transformers and BERT-based pre-training within NLP has motivated us to support a more detailed understanding of these models.

\subsection{Model Interpretability}
\label{subsec:ref_interp}

The successes of BERT have led to numerous works in attempting to understand \emph{why} the model -- specifically the Transformer model and its pre-training objectives -- performs so well~\cite{rogers2020primer}. Existing works are largely targeted at understanding the representations learned during pre-training. Specifically, recent works~\cite{clark2019does,reif2019visualizing} highlight how attention heads, across different layers, capture various dependency relations, e.g. prepositions and coreferent mentions. Lin et al.~\cite{lina2019open} show how learned embeddings capture subject nouns and main auxiliaries, while attention captures subject-verb agreement and anaphora relations. Brunner et al.~\cite{Brunner2020On} demonstrate that word embeddings in earlier layers tend to retain their identity, while deeper layers represent aggregated, abstract information. Other works have studied the differences in BERT between pre-trained models and their fine-tuned counterparts. For instance, Hao et al.~\cite{hao2019visualizing} visually illustrate the loss landscapes produced during fine-tuning, while other works inspect how syntactic relations change by inspecting BERT's attention patterns and embeddings~\cite{kovaleva2019revealing,van2019does}.

These methods can highlight relevant linguistic phenomena as high-level summaries over a given dataset, typically through a pre-defined probing task (e.g. classifying dependency relations). However, there are two main limitations with these methods. First, they do not permit \emph{local explanations} of individual instances, e.g. how can we understand the changes made from pre-training to fine-tuning on a given sentence pair? Secondly, the focus on individual attention heads, or layers, does not permit \emph{global interpretability} of the model: the relationship between attention across multiple layers. These aspects of explainability are key for good user experiences in explainable AI systems~\cite{hohman2019gamut,liao2020questioning} and reflect objectives that we target in our visualization design.

\subsection{Visual Analytics for Interpreting Models}
\label{subsec:ref_va}

Within the visual analytics community, significant research has been devoted to analyzing and interpreting machine learning models, in particular deep learning methods, please see the survey by Hohman et al.~\cite{hohman2018visual}. Although much work has been devoted to understanding convolutional neural networks for image classification~\cite{liu2016towards,pezzotti2017deepeyes,hohman2019s}, here we discuss techniques most relevant to our method, namely techniques for visually analyzing NLP models. In the context of RNNs, as previously discussed in Section~\ref{subsec:ref_lm}, Strobelt et al.~\cite{strobelt2017lstmvis} visualized hidden states as line marks plotted over a sequence, one per dimension, while Ming et al.~\cite{ming2017understanding} seek to group hidden state activations via clustering. Other works are more task-specific, e.g. interactively exploring machine translation~\cite{strobelt2018s} and inspecting classification decisions in natural language inference tasks~\cite{liu2019roberta}, while Cashman et al.~\cite{cashman2019rnnbow} focus on understanding the training of recurrent networks through visualizing network gradients, and Gehrmann et al.~\cite{gehrmann2019visual} support fine-grained model editing for abstractive summarization.

Other works have begun to address the visualization of attention mechanisms in language models. Some works have visually inspected attention for RNNs~\cite{strobelt2018s,liu2018nlize}, in particular, Choi et al.~\cite{choi2019aila} showed how visually encoding attention in an RNN-based sentiment classification task can help humans more efficiently and effectively annotate data. More recent work has started to consider the visualization of Transformer models~\cite{vig2019multiscale,hoover2019exbert,park2019sanvis}. These works are largely focused on visually analyzing individual attention heads~\cite{vig2019multiscale,hoover2019exbert}, or alternatively the aggregation of attention heads in a given layer~\cite{park2019sanvis}, and provide support for querying sentences based on selected head embeddings~\cite{hoover2019exbert}. Rather than individually inspect attention heads or layers, our work aims to provide a more holistic view of self-attention in Transformers so that it is possible to visually analyze and compare models in terms of a particular downstream sentence understanding task.

\subsection{Comparative Visualization}
\label{subsec:ref_compare}

A major emphasis of our work is on the visual comparison of language models and in particular, the comparison of graphs that result from their self-attention mechanisms. Visual comparison has been extensively studied within the visualization community, please see Gleicher~\cite{gleicher2017considerations} for an overview. For visually comparing graphs, existing works have visually depicted similarities via graph merging and using color encodings~\cite{andrews2009visual,koop2013visual}, while a larger design space -- heatmaps, grouped bars node-link diagrams -- has also been studied~\cite{alper2013weighted}. Our design is inspired by these works, yet in our scenario the graphs from models are dynamic, updated in response to user interactions. Other works have considered the visual comparison of deep networks, specifically convolutional networks~\cite{zeng2017cnncomparator} and recurrent models~\cite{murugesan2019deepcompare}. However, these works are largely focused on comparing model performance, whereas our method is aimed at comparing how models reason over a given input.

%% file: tasks.tex

\begin{figure}[!t]
    \centering
    \includegraphics[width=0.96\linewidth]{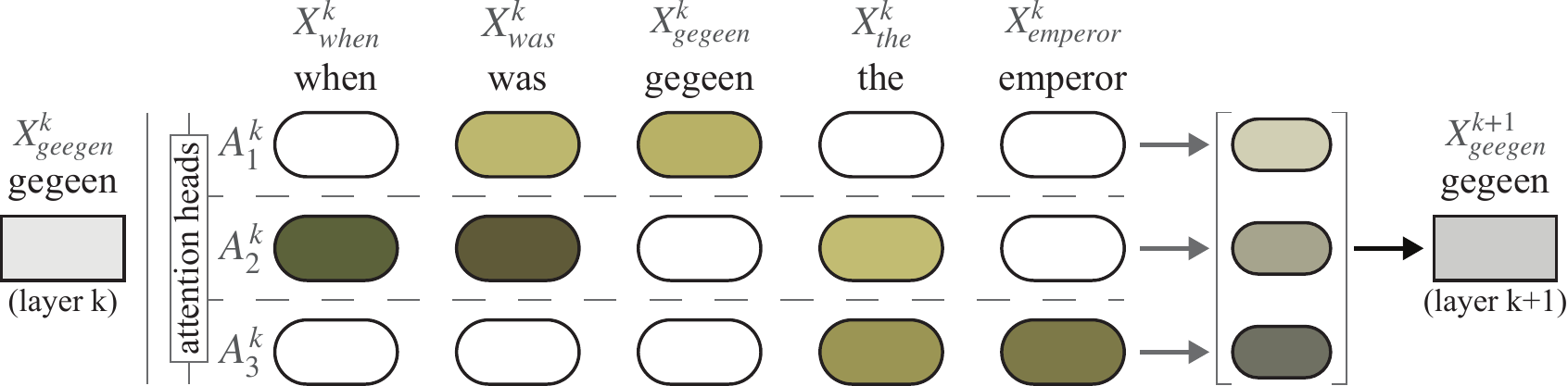}
    \caption{We illustrate the computation involved in self-attention for the sequence ``when was gegeen the emperor''. Here the word ``gegeen'' attends to the rest of the sequence in independent attention heads, producing a separate embedding for each head. These are combined to produce a new embedding for ``gegeen'' at the next layer.}
    \label{fig:att_illustrate}
\end{figure}

In this section we provide details on the Transformer model, BERT, and the set of tasks we aim to address through our visualization design.

\subsection{A Summary of BERT}
\label{subsec:SummaryBERT}

Our model of study is BERT~\cite{devlin2019bert}, a Transformer model~\cite{vaswani2017attention} pre-trained to solve certain language modeling objectives. For clarity of notation, herein we refer to words as tokens, in order to encompass a larger set of inputs, e.g. punctuation, specialized symbols, as well as word pieces~\cite{wu2016google}. The Transformer model relies on a notion of \emph{self-attention} to compute \emph{contextualized} token embeddings for a provided sequence of tokens. Context refers to a token's position, and other tokens that surround it, in a given sequence. More specifically, assume that we are provided a sequence of tokens $s = (w_1, w_2, \cdots , w_n)$. The Transformer applies a stack of self-attention layers to a sequence, where the input of one layer is the output embedding of the previous. Assume that we are considering a layer $l$, we have an embedding vector $\mb{x}^l_w \in \mathbb{R}^d$ for $w \in s$, $d$ is the embedding dimension, and we denote the sequence matrix as $X^l \in \mathbb{R}^{n \times d}$, e.g. each row corresponds to a token's embedding. To obtain a new contextualized embedding in the next layer $X^{l+1}$, the Transformer (1) performs self-attention, one per attention head, (2) derives a new vector from each attention head, and (3) combines the per-head vectors into a single vector.

\textbf{1. Self-attention:} For a given \emph{attention head} indexed by $j$, a dot product-based \emph{attention matrix} is formed,
		\begin{equation}
			A_j^l = \text{softmax}\left(\frac{1}{\sqrt{d'}} X^l Q_j^l \left(X^l K_j^l \right)^{\intercal}\right),
		\end{equation}
		where $Q_j^l, K_j^l \in \mathbb{R}^{d \times d'}$ are projection matrices from a $d$-dimensional space to a $d'$-dimensional space where $d' < d$, and the softmax function is applied row-wise; each row sums to 1 and its entries are nonnegative.
		
\textbf{2. Attention-based Vectors:} A new embedding vector is formed for each token, and each attention head,
		\begin{equation}
			X_j^l = A_j^l X^l V_j^l,
		\end{equation}
		where $V_j^l \in \mathbb{R}^{d \times d'}$ is a projection matrix, similar to $Q$ and $K$. Given the properties of $A_j^l$, each token's output embedding is formed as a convex combination over the input token embeddings. Thus, an entry at row $a$ and column $b$ in $A_j^l$ may be interpreted as how \emph{important} token $w_b$ is to token $w_a$. Fig.~\ref{fig:att_illustrate} illustrates this computation, where each row corresponds to a head's self-attention for the given token ``gegeen'', depicting weights over tokens in the sentence.
		
\textbf{3. Vector Aggregation:} The embedding vectors are then concatenated over all heads: $X_{\text{cat}}^l = \text{cat}\left[X_1^l , X_2^l , \cdots , X_h^l\right] W^l$, with $W^l \in \mathbb{R}^{d \times d}$. Note that if one token attends to another token over multiple attention heads, then it will have more influence on the output via this concatenation. This is depicted in Fig.~\ref{fig:att_illustrate} (right), where we can see how ``gegeen'' aggregates different types of context, and ``emperor'' has less overall influence. After concatenation, layer normalization~\cite{ba2016layer} and residual connections~\cite{he2016deep} are applied to the original and concatenated vectors, and last, a multi-layer perceptron (MLP) is applied to this result, followed by another application of layer normalization and residual connections.

BERT~\cite{devlin2019bert} uses the Transformer model to learn contextualized embeddings that solve two unsupervised \emph{pre-training} tasks. First, a masked token prediction task randomly replaces tokens in a sequence with a unique ``mask'' token, and the model learns to predict a probability distribution that assigns high likelihood to the original tokens. Second, a next sentence prediction task aims to distinguish sentence pairs that are randomly sampled from sentence pairs that form a single contiguous sequence. The collection of projection matrices at all layers and attention heads, as well as MLPs, form the set of weights of the model to be learned during pre-training. Once completed, a downstream supervised sentence understanding task, e.g. question answer inference, may then be trained by \emph{fine-tuning} the weights of the model. Typically, the process takes a reserved classification ([CLS]) token's embedding at the very last layer, which we denote $\mb{x}_{CLS}^L$, and projects the embedding to a set of classification scores for prediction.

\begin{figure}[!t]
    \centering
    \includegraphics[width=.9\linewidth]{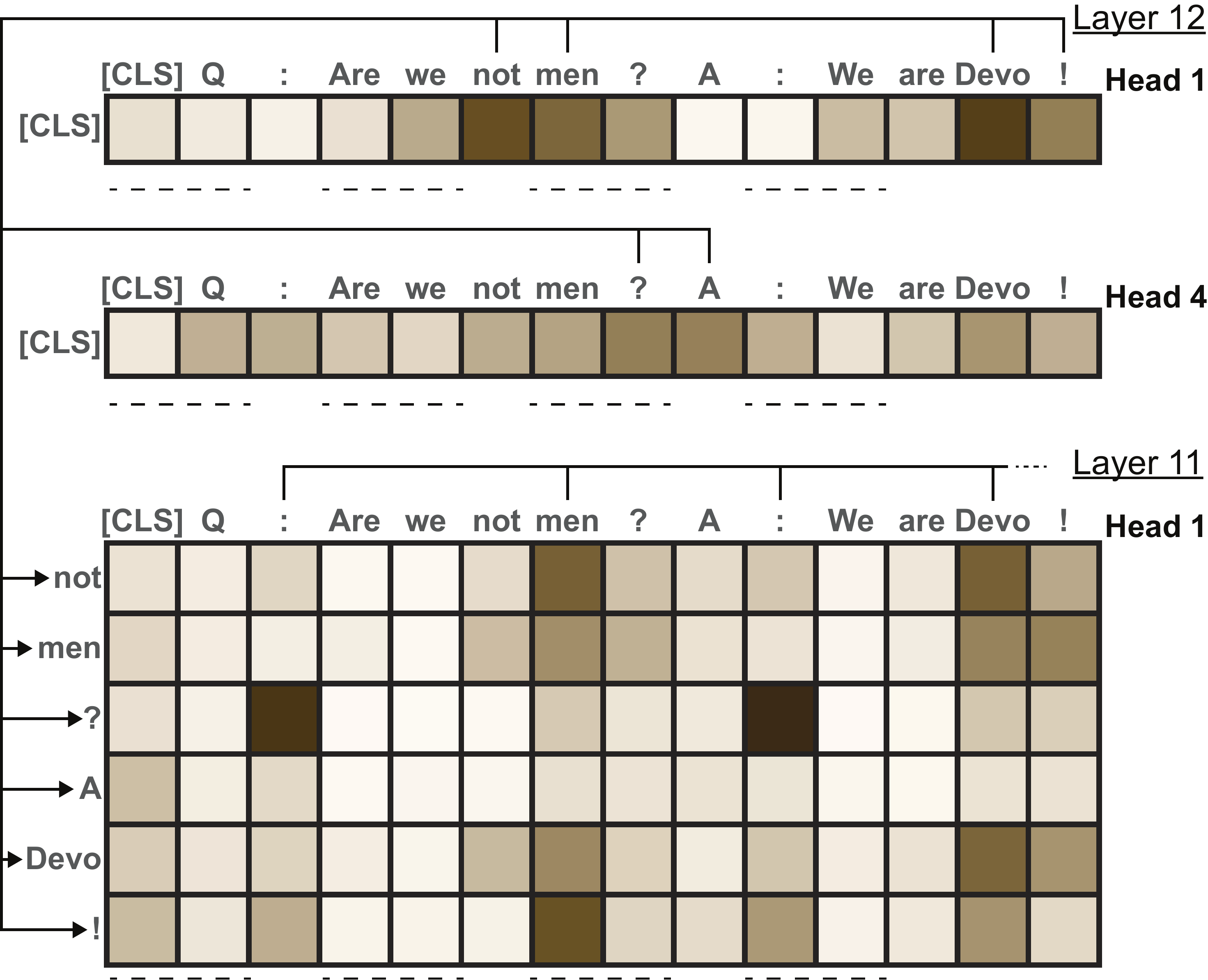}
    \caption{The Attention Graph is built by adding edges between tokens that contain high attention. The [CLS] token (top) is the sole node for the last layer (12), and all extracted tokens (``not'', ``men'', ``?'', ``A'' ``Devo'', ``!'') become new nodes for the prior layer (bottom).}
    \label{fig:attgraph}
\end{figure}

\subsection{Objectives and Tasks}
\label{subsec:tasks}

The main focus of our work is in understanding how BERT performs classification when fine-tuned for NLP tasks, and thus, our starting point is the classification embedding $\mb{x}_{CLS}^L$. Yet, the complexity of BERT poses challenges for gaining insight, with components ranging from attention to contextualized embeddings to various learned transformations. In this work, we have chosen to study the self-attention mechanism, specifically the set of matrices $A_j^l$ for all attention heads $j$ and layers $l$, and how self attention organizes to form $\mb{x}_{CLS}^L$. We view these matrices as key elements in understanding \emph{information flows} in the Transformer model, e.g. how does one token influence another token across the layers of the network? Thus, in contrast to prior work that is more exploratory regarding self-attention~\cite{vig2019multiscale,hoover2019exbert}, our work seeks to analyze attention to help \emph{explain} the classification decisions made by BERT. We note that self-attention is not the only way to quantify classification influence, as gradient-based attribution schemes are also commonly employed~\cite{serrano2019attention}. However, for pre-trained models, task-specific parameters attached to the [CLS] token have yet to be updated as part of training, and thus gradients are not particularly meaningful. Simple model architectures for fine-tuning~\cite{devlin2019bert} therefore suggest $\mb{x}_{CLS}^L$, computed through self-attention, will remain important for prediction.

We thus identify two main objectives that we aim to address:
\begin{enumerate}
	\item \textbf{Understand how self-attention informs classification (O1)}. We would like to understand how the model makes decisions via the words that it attends to, starting from the classification token at the very last layer and going backwards in layers.
	\item \textbf{Understand the refinement of self-attention due to fine-tuning (O2)}. We would like to assess what was learned by the model when fine-tuned for a specific task, in order to confirm that the model is learning relevant information in solving the task.
\end{enumerate}
Note that \textbf{(O2)} is dependent on \textbf{(O1)}: fine-tuning is tied with classification, thus it is challenging to identify changes in self-attention without an understanding of how self-attention is used to form classifications.

We address these objectives via the following tasks:
\begin{enumerate}
	\item \textbf{Trace and query self-attention throughout the model (T1)}. The complexity of self-attention requires user interactions that support the selection of tokens and attention heads over different layers in the model to understand how attention propagates forward (deeper in layers), as well as how attention dependencies form from shallower layers \textbf{(O1)}.
	\item \textbf{Discover attention functionality over layers and attention heads (T2)}. An understanding of the model requires comprehending self-attention in aggregate and individual heads, discovered via user queries over interpretable units, e.g. input tokens \textbf{(O1)}.
	\item \textbf{Compare self-attention of pre-trained and fine-tuned models (T3)}. To understand fine-tuning, it is necessary to assess similarities and differences between pre-trained and fine-tuned models. This should allow for (\textbf{a}) detailed comparisons, e.g. locating shared and distinct attention heads, and (\textbf{b}) global comparisons, e.g. tracing differences in attention flows across layers \textbf{(O2)}.
\end{enumerate}

%% file: design.tex
Our visualization design is informed by the tasks identified in the previous section. Here we first discuss the information that we extract for our visualization, followed by a discussion of our design.

\begin{figure}[!t]
    \centering
    \includegraphics[width=1\linewidth]{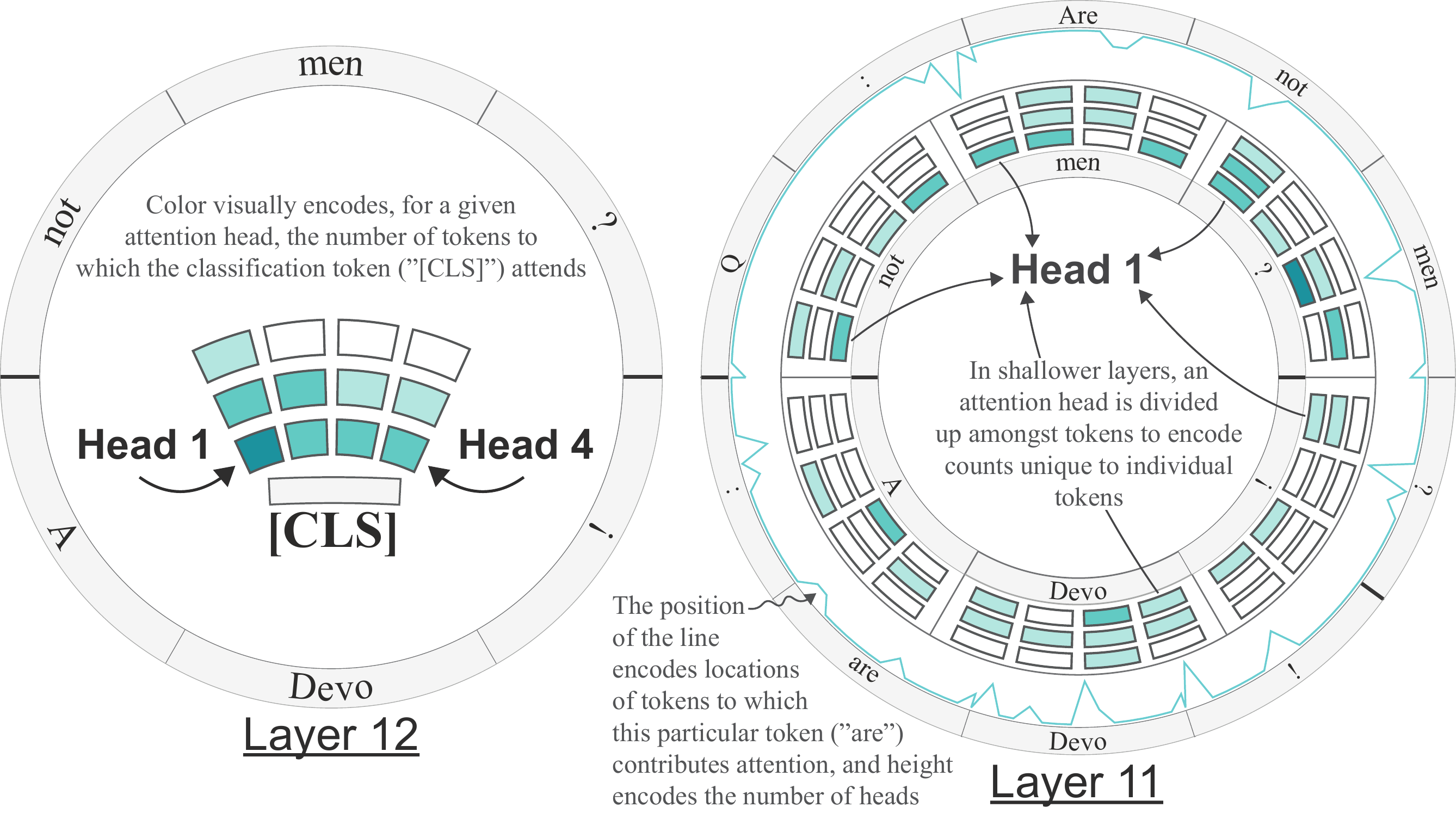}
    \caption{At the last layer, the tokens that [CLS] depends on (c.f. Fig.~\ref{fig:attgraph}) are encoded in a ring (left), where the attention heads for [CLS] are shown in the center. This design is carried over to previous layers (right), where per-token attention information is depicted, along with the positions of attended tokens.}
    \label{fig:encoding}
\end{figure}

\subsection{Attention Graph}
\label{subsec:att_graph}

As discussed in Sec.~\ref{subsec:tasks}, our main focus is on the analysis of self-attention. However, analyzing individual attention matrices, or even aggregate attention over a single layer~\cite{park2019sanvis}, might not capture all of the information necessary to understand BERT's classification. Instead, in this work, we aim for a \emph{global} view of attention -- over all heads and layers -- as a means of understanding how attention flows from tokens at arbitrary layers down to the classification ([CLS]) token. To this end, we would like to capture the prominent dependencies between tokens when computing contextualized embeddings from one layer to another. Specifically, for an attention matrix $A_j^l$, we construct a bipartite graph $G_j^l$, where an edge between tokens $w_a$ and $w_b$ is formed if $A_j^l(a,b) > \tau$ for a threshold $\tau$, a parameter that we allow the user to interactively modify. In this case, $w_b$ corresponds to this token's embedding at layer $l$, and such an edge indicates $w_b$'s strong influence on the embedding for $w_a$ at layer $l+1$. The graph is constructed iteratively, starting from the [CLS] token at the last layer, and working backwards in layers, please see Fig.~\ref{fig:attgraph}.

Considering all bipartite graphs across all layers permits us to trace dependencies between tokens. However, what remains is a way to define such a dependency. Specifically, for tokens $w_b$ and $w_a$, and subsequent layers $l$ and $l+1$, respectively, we consider two options:
\begin{enumerate}
	\item $w_b$ is dependent on $w_a$ if there exists \emph{some} attention head $j$ at layer $l$ that connects the tokens in its graph $G_j^l$. This scheme is enabled by default in our design.
	\item A dependency only exists between two tokens if it passes through a \emph{subset} of attention heads. This is enacted via user interactions.
\end{enumerate}
Given one, or a combination, of methods for constructing dependencies, we may then form a larger graph between tokens over all layers, where edges only exist across adjacent layers. We call this graph the \emph{Attention Graph}, and it is the primary object that we analyze. The Attention Graph enables us to trace information flows between tokens, across layers, and amongst heads, addressing (\textbf{T1}). For instance, a token's embedding at layer $l$ is unlikely to influence another token's embedding at layer $m$ ($m > l$) if no path exists between them. If a single path does exist, but it is a long path (e.g. over multiple layers), then the token's influence will be minimal, as it's identity is likely to be lost between layers~\cite{Brunner2020On}. If multiple paths exist between the tokens, then this represents strong evidence of influence.

\subsection{Attention Overview}
\label{subsec:att_summary}

Our design, first, provides an overview of a model's self-attention mechanism, depicting a notion of \emph{classification influence} for each token in the sentence at a given layer. To compute influence for a given token $w$ at the last layer $L$, we record how many heads were used in attending to the [CLS] token, denoted as $c_L(w,[CLS])$ for token $w$. For the previous layer $L-1$ and token $w$, we gather attended tokens at layer $L$, denoted $W_L$, and record the number of counts over these tokens:
\begin{equation}
    c_{L-1}(w) = \sum_{w' \in W_L} c_{L-1}(w,w'),
\end{equation}
where $c_{L-1}(w,w')$ indicates the number of heads used from $w$ to attend to $w'$. We iteratively apply this scheme, working backwards in layers, to define a count for any token $w$ at layer $l$ via $c_l(w)$. We then compute an influence score that summarizes all layers, from a given layer $l$:
\begin{equation}
I_{l}(w) = \frac{1}{L-l+1} \sum_{l'=l}^{L} \alpha^{L-l'} c_{l'}(w),
\end{equation}
which averages the scores over all layers, applying an exponential decay to earlier layers as contextualized embeddings at these layers are unlikely to be as significant as embeddings at later layers, where we set $\alpha= 0.5$. We find that different values of $\alpha$ do not impact the relative comparisons of models.

In our Sentence view, we visually encode a token $w$'s influence, given a user-selected layer $l$, by taking the ceiling of $I_{l}(w)$ and mapping this as a 5-circle rating directly above the token (see Fig.~\ref{fig:teaser}(a)). We clamp all influence scores to 5, as certain tokens, e.g. punctuation, can attend to a large number of tokens. Further, for model comparison, we map the circles to a dark orange color when both models share a certain amount of influence and use distinct colors when influence scores differ, e.g. if $I_{l}(w) = 2$ for the pre-trained model and $I_{l}(w) = 4$ for the fine-tuned model, then the first two circles will be orange, and the last two circles will be purple. This provides an at-a-glance comparison between the models in their self-attention influence on the classification score, addressing the overview aspect of task (\textbf{T3-b}), and can be used to identify tokens of interest for more detailed analysis, discussed next.

\subsection{Attention Flows Design}

The Attention Flow view uses the Attention Graph to depict self-attention across multiple layers and multiple attention heads, for a given sentence, as shown in Fig.~\ref{fig:teaser}(c). Our visualization design depicts various aspects of the model: token dependencies, attention head information, and the capacity to visually compare models.

\subsubsection{Token Rings}

For our central view, we deploy a radial layout, where each ring of tokens corresponds to a given layer -- please see Fig.~\ref{fig:encoding}. The innermost ring includes only the [CLS] token, as this is the deepest layer before classification. The preceding ring contains tokens from the prior layer that have some influence on the embedding of the [CLS] token, namely, there exists \emph{some} attention head that connects a given token and [CLS], as discussed in Sec.~\ref{subsec:att_graph}. This process is repeated, aggregating the previous layer's tokens that have some influence on the current layer's tokens, allowing us to trace token influence throughout the model (\textbf{T1}).

This design is intended to effectively utilize space in showing dependencies between tokens across multiple layers of the Transformer model. In practice, for the [CLS] token at the last layer (center of view), the number of tokens that it depends on from the previous layer is relatively small and thus, may be visually encoded in an annulus with small radii. The number of tokens in preceding layers tends to grow by a bounded amount, in practice by at most 5-10 tokens, and thus, a radial view scales well with the number of tokens in each layer. Each ring of tokens aims to capture the gist of the sentence at different levels of detail, depending on the layer, e.g. in Fig.~\ref{fig:teaser} the penultimate layer contains the tokens necessary to answer the question. To provide context with respect to the original sentence, we depict the positions of tokens found in each layer in an auxiliary radial view (c.f. Fig.~\ref{fig:teaser}(b)), where each token's position in the original sentence is encoded via angle. Further, to increase readability of tokens, we show as many tokens right side up as possible by having both sentences start from the same angle of the radial layout, with the first sentence going clockwise and the second sentence going counterclockwise until their meeting point (see Fig.~\ref{fig:teaser}), depicted as a thick black tick at each layer.

\begin{figure}[t]
    \centering
    \includegraphics[width=.78\linewidth]{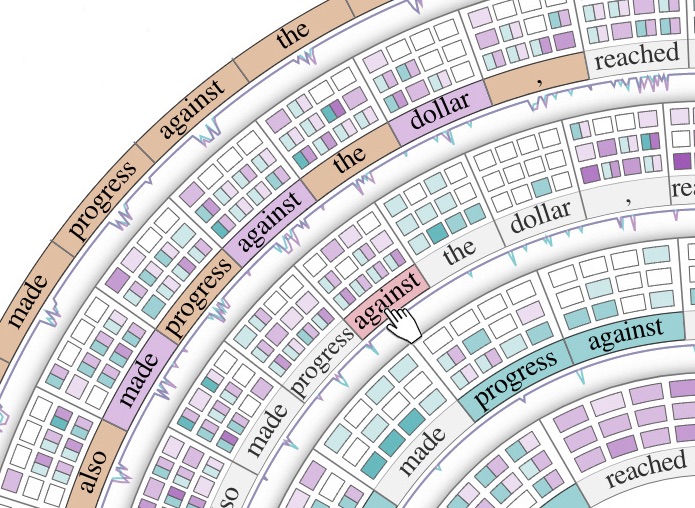}
    \caption{Our design allows the user to compare model attention: here the user clicks on "against", a token that shares influence between models. Inspecting its dependent tokens in previous layers, we observe commonalities (in orange) and tokens unique to the fine-tuned model (purple), e.g. "dollar". Attention heads for tokens are, further, split between models where appropriate.}
    \label{fig:hovering_2models}
\end{figure}

\subsubsection{Attention Heads}

The tokens within each ring represent dependencies that exist regarding some attention head. To provide more detailed information, we visually encode attention heads in two different manners: with respect to \emph{rows} of a head's attention matrix and with respect to its \emph{columns}. Recall that rows correspond to tokens that are performing attention and reflect what is \emph{output}. Thus, adjacent to each token, we visually encode its attention heads as a set of 12 small glyphs (one for each head), where the color encodes the number of tokens a head attends to in the previous layer, shown in Fig.~\ref{fig:encoding}. We wrap the 12 head glyphs across 3 rows to optimally use space in the design. This design permits a more detailed assessment of token dependencies, e.g. determining if a set of heads collectively attend to a large number of tokens (\textbf{T2}).

\begin{figure}[t]
\centering
\includegraphics[width=1\linewidth, height=.5\linewidth]{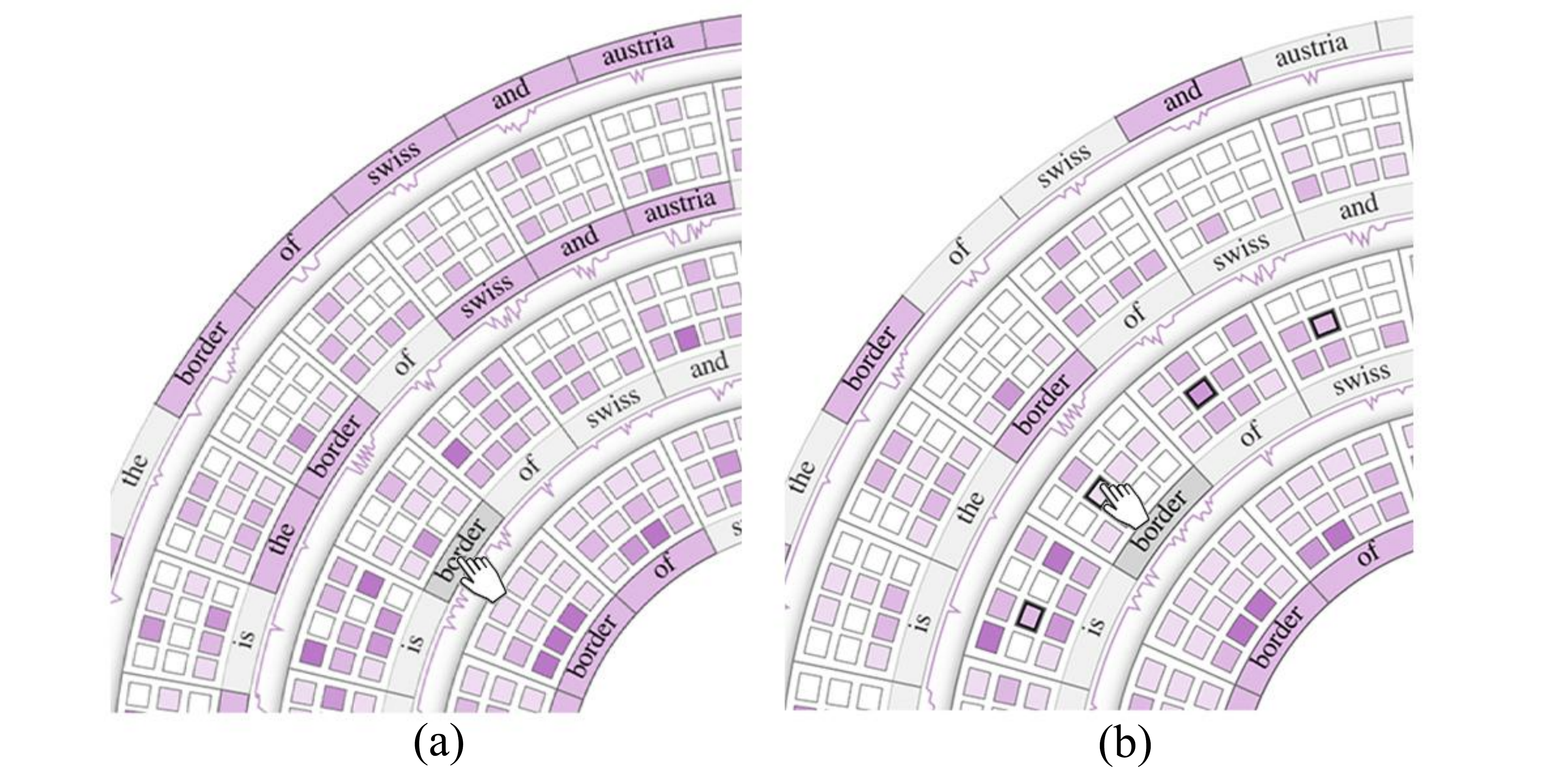}
\caption{Display of attention propagation while hovering over a token: (a) shows hovering over a single token and (b) shows hovering over a head glyph for propagation through only that head.}
\label{fig:Hovering}
\end{figure}

Columns correspond to tokens that are being attended to in the \emph{input}. We visually encode this information by mapping the relative positions of tokens at layer $l+1$ to sparklines under each dependent token at layer $l$, shown in Fig.~\ref{fig:encoding} (right). The location of the peak indicates whether the dependency occurs earlier (peaks left of center) or later in the sequence (peaks right of center), with a token attending to itself if the peak is in the center. The height of the peak encodes the number of attentive heads, clamped to 3, involved in the dependencies. This design allows the user to quickly assess whether self-attention is localized (near the center) or more global (a sparkline that is uniformly distributed).

\subsubsection{Model Comparison}

Our design supports a combined model display to facilitate comparison, please see Fig.~\ref{fig:hovering_2models} for an illustration. We take the union of tokens attended to by each model and visually encode a token's detailed attention heads with colors that are model-specific, and shared between models (\textbf{T3-a}). If a head only has significant attention in BERT, we color the head glyph turquoise. If a head only has significant attention in the fine-tuned model, we color the head glyph purple. If a head has significant attention in the Attention Graphs of both models, we split the head glyph in half to indicated shared attention (Fig.~\ref{fig:hovering_2models}). This color scheme is carried over to the Sentence Context view to create homogeneity and facilitate quick model comparisons, specifically, the parts of the sentence included in each Attention Graph.

\begin{figure}[t]
    \centering
    \includegraphics[width=0.5\linewidth]{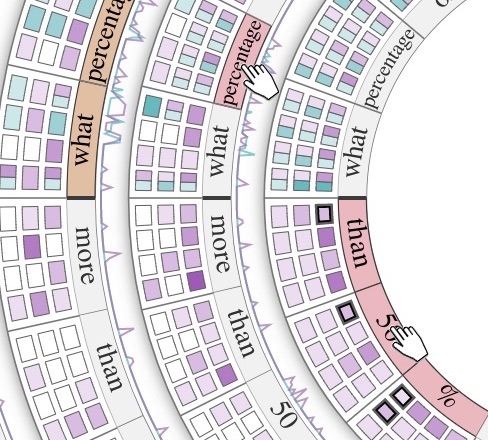}
    \caption{Head filtering: a user selects token ``percentage" at a specified layer denoted $L$, and tokens ``than", ``50", \& ``\%" via brushing in the subsequent layer $L+1$. For each selected token in layer $L+1$, heads connecting to selected tokens in layer $L$ have their glyphs highlighted.}
    \label{fig:head_filtering}
\end{figure}

\subsection{Interacting with Attention}

Our design supports a number of interactions to help users gain better insight into a model or models.

\subsubsection{Exploring Attention Subgraphs}

We allow the user to view subgraphs by selecting individual tokens at arbitrary layers. For a given selected token, we highlight tokens at subsequent layers that depend on the selection, as well as all tokens at prior layers that the selection depends on (\textbf{T1}), please see Fig.~\ref{fig:Hovering}(a) for an example. A pair of tokens are considered dependent if there exists a path between them in the Attention Graph. By default, edges in the graph are formed if any attention head exists connecting a pair of tokens; for investigating specific heads, a user may select a head glyph for a given token, which limits connectivity to that specific head and given layer, shown in Fig.~\ref{fig:Hovering}(b). At all other layers, all heads are still used for determining paths.

Head selection is enacted via mouse hovering, as well as clicking, in order to freeze the current selection. Further, to select token sequences, we support brushing within layers, where we highlight the intersection of the attention graphs for the brushed tokens. The intersection is performed at each layer before continuing traversal at the next layer. This can be used to assess whether contiguous text phrases (e.g. named entities) have similar attention dependencies.

\begin{figure}[t]
    \centering
    \includegraphics[width=0.9\linewidth]{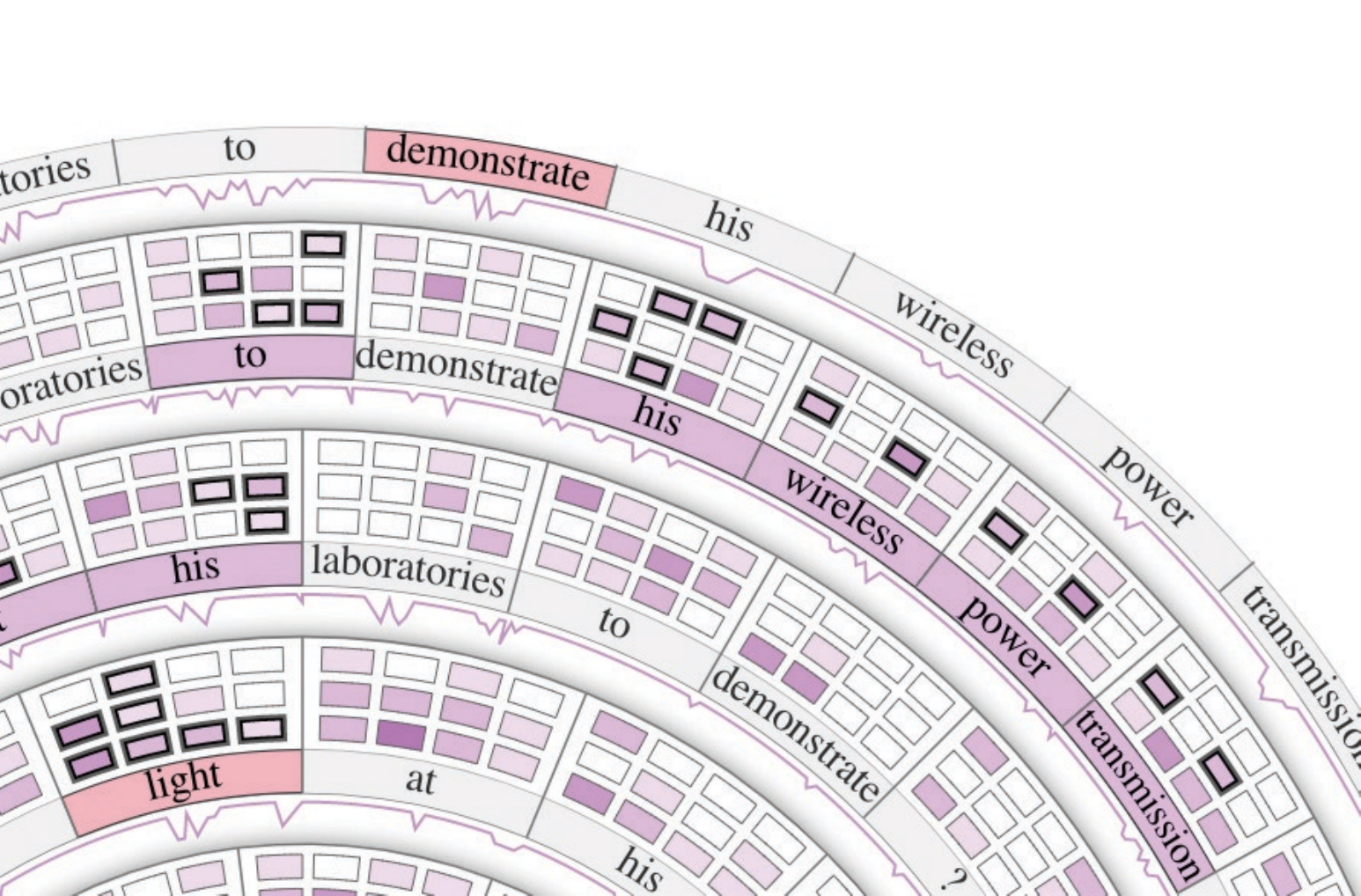}
    \caption{Selecting two tokens in non-adjacent layers highlights all tokens -- and attention heads -- through which attention flows, in this example from ``demonstrate'' to ``light''. We find tokens in the phrase ``wireless power transmission'' all share the same heads in the graph traversal.}
    \label{fig:intermediate_headfiltering}
\end{figure}

In the combined view, we offer a consistent form of hovering as with the single model view: hovering over a token will still highlight the attention traversal using the attention heads of that token. We perform traversals for the present attention heads of each model for the hovered token, union the resulting tokens, and highlight all tokens. We treat the attention graph of each model as separate entities, so the traversal for each model is computed separately (see Fig.~\ref{fig:hovering_2models}). Thus, hovering over a token with attention heads from only one model will result in identical results to the single model view. This can be used to understand global similarities/differences between models (\textbf{T3-b}), e.g. for a token in a shallow layer, what tokens are dependent on it in deeper layers, and which models contain such dependencies. For a coherent user experience, we keep the color scheme uniform with the Sentence Context view and the head glyph coloring (pre-trained -- turquoise, fine-tuned -- purple, overlap -- orange).

Hovering over head glyphs still limits traversals to that attention head, except the split head glyphs now consist of two separate hover targets. We allow users to both select individual models by hovering over the appropriate glyph, as well as to depict model comparisons via holding an appropriate keybinding when hovering over either glyph.

\begin{figure}[t]
    \centering
    \includegraphics[width=1.0\linewidth]{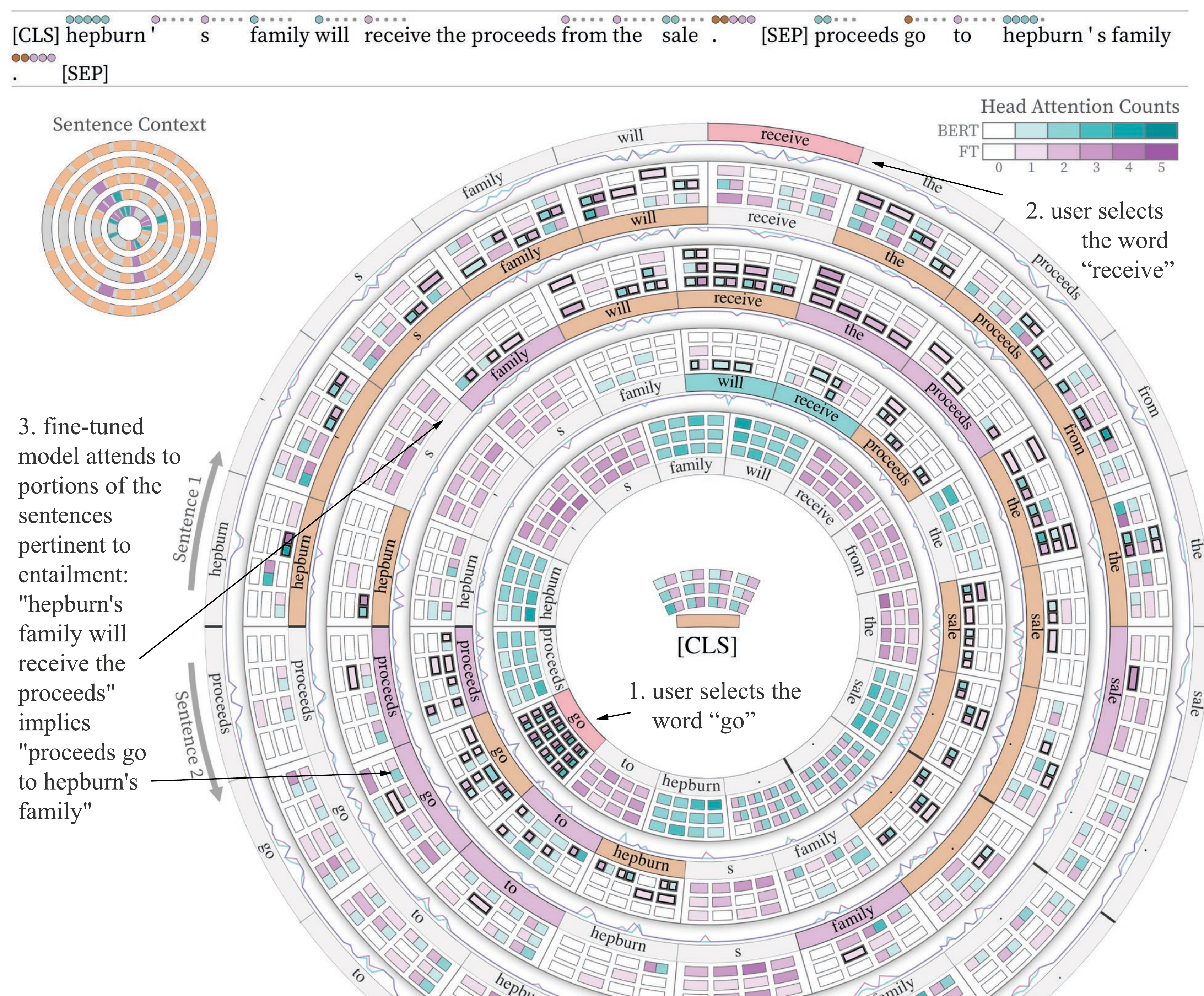}
    \caption{We showcase our interface for the task of recognizing textual entailment. By selecting the words ``receive'' and ``go'', we highlight how attention flows between the two sentences, aggregating information that is necessary to determine that ``will receive'' implies ``proceeds go to'', regarding ``hepburn's family''.}
    \label{fig:rte_use_case}
\end{figure}

\subsubsection{Querying Head Functionality}

In order to provide a more intuitive means of assessing head functionality, we allow the user to select tokens \emph{across} layers, highlighting the set of heads through which attention flows (\textbf{T2}).
Further, for tokens that are not in adjacent layers, we highlight intermediate tokens on the path between the selected tokens, along with their respective attention heads, as shown in Fig.~\ref{fig:intermediate_headfiltering}. If multiple tokens are selected in either layer, the intersection path, or intermediate tokens that are attended to by all selected tokens, is shown.

In the combined model view, selecting tokens in different layers behaves consistently with the single model view. We traverse the attention graph for the BERT model and fine-tuned model separately, searching for paths through attention heads connecting the selected tokens with tokens in the path highlighted, with our consistent color scale according to which model's attention graph from which it was extracted. Paths are not required to join all selected tokens but only those included as nodes in their respective attention graph.

%% file: experiments.tex

We showcase Attention Flows through a set of use cases and user feedback. Our focus is on highlighting model differences in BERT, e.g. what did fine-tuning learn for a particular task? To this end, our method is used on a benchmark set of supervised NLP tasks.

Our experiments are conducted in a similar manner to~\cite{devlin2019bert}. We use the pre-trained, uncased BERT\textsubscript{BASE} model, a Transformer model with 12 layers and 12 attention heads per layer. Fine-tuning BERT is straightforward as task specific inputs to BERT, e.g. hypothesis-premise pairs in entailment, are analogous to sentence pairs during pre-training (see Sec.~\ref{subsec:ref_lm}). For fine-tuning, given a sentence understanding task, we introduce a linear layer for classification, which takes as input the [CLS] token embedding from the very last layer of the model. We represent the input sequence as a sequence pair, separated by a reserved [SEP] token, and use the final hidden vector $\mb{x}_{CLS}^{12} \in\mathbb{R}^{d}$, corresponding to a reserved [CLS] token, as an aggregate representation of the sentence~\cite{devlin2019bert}. We compute a standard classification loss (cross entropy) using the introduced linear transformation $W\in\mathbb{R}^{k \times d}$, where $k$ is the number of labels, introduced in the fine-tuning classification layer. During training, we update both $W$ and the Transformer's model weights, via stochastic gradient descent using the Adam optimizer~\cite{kingma2014adam}. We obtain comparable development accuracy on all GLUE tasks to those reported in Devlin et al.~\cite{devlin2019bert}, specifically, 91\% for QNLI, 86.5\% for MPRC, and 68.6\% for RTE, tasks that we detail next.

\begin{figure}[t]
    \centering
    \includegraphics[width=1.0\linewidth]{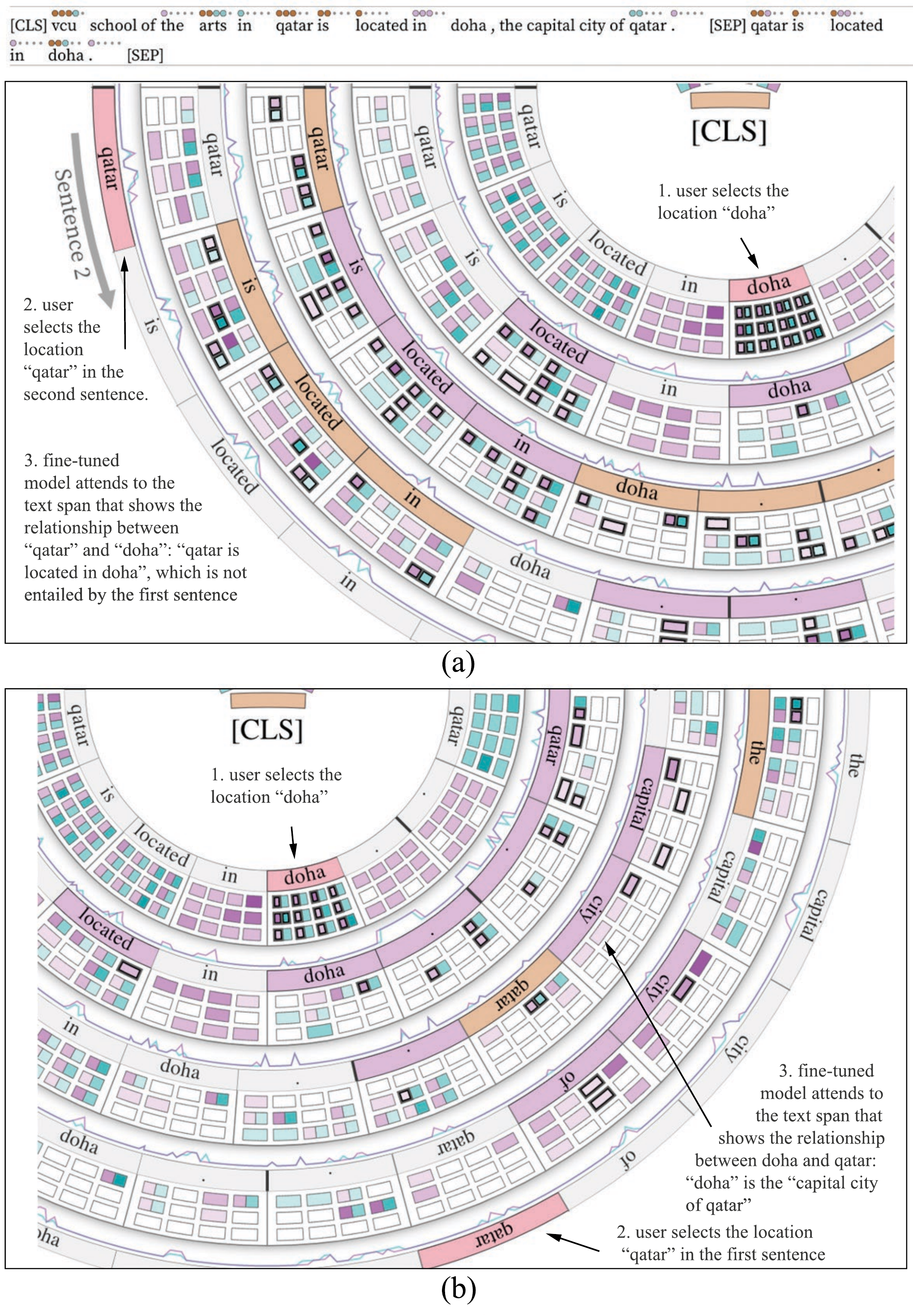}
    \caption{In this entailment example, the first sentence does not entail the second. Here, the user selects ``qatar'' in the entailed (a) and entailing (b) sentence to understand how attention flows from these words to ``doha''. The fine-tuned model attends to the past participle ``located in'' (a) that comprises the hypothesis, while the entailing (b) sentence shows attention that describes doha as the ``capital city'' of qatar.}
    \label{fig:rte2}
\end{figure}

\subsection{Datasets}
The General Language Understanding Evaluation (GLUE)~\cite{wang2018glue} benchmark is a collection of sentence understanding tasks designed to evaluate the performance of NLP models. We evaluate our tool by investigating self-attention on three of the nine sentence-pair, binary classification-based, GLUE tasks: RTE, MRPC, QNLI.

\begin{figure*}[!t]
    \subfloat[\label{subfig:qnli1}]{
         \includegraphics[width=0.518\linewidth]{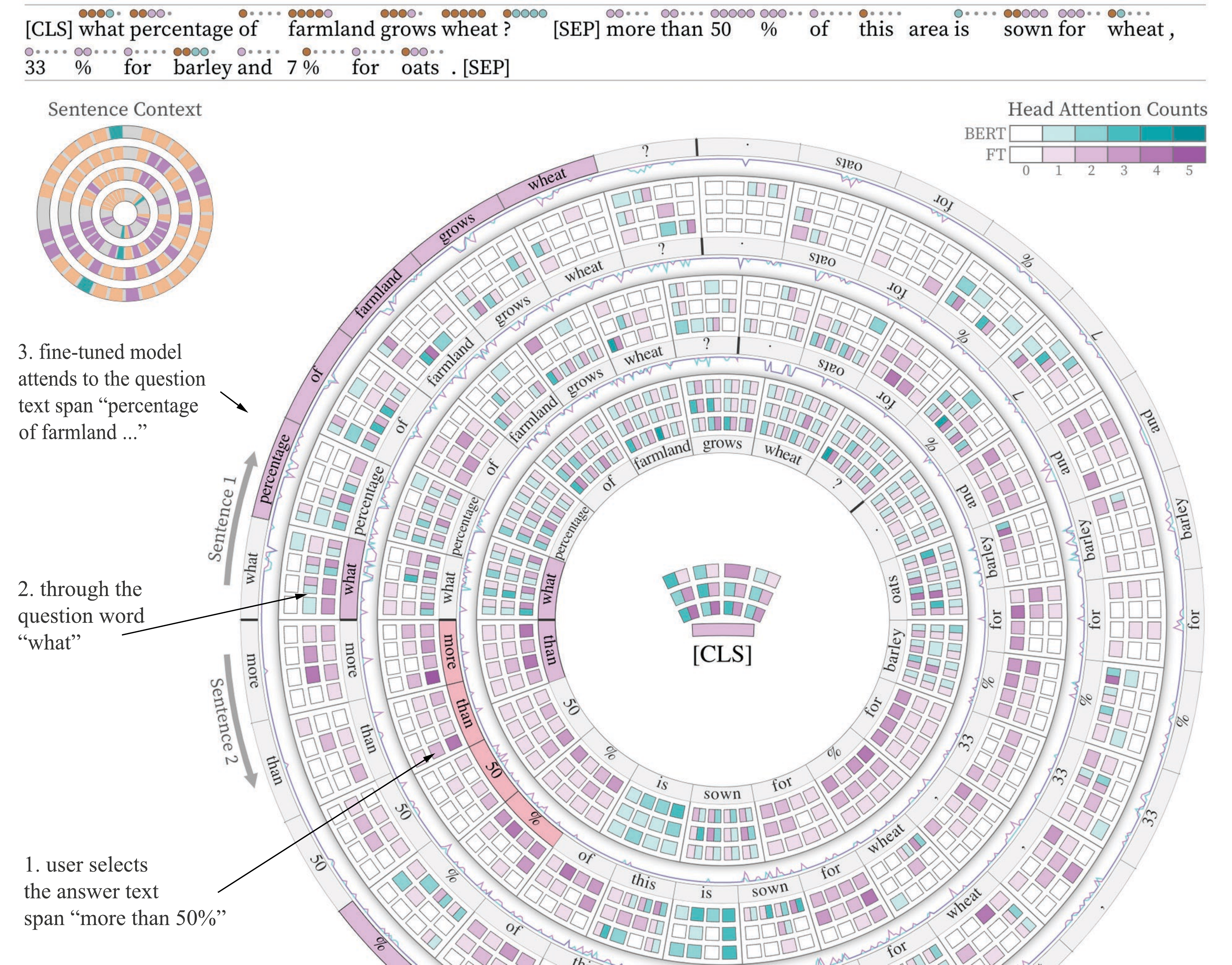}
	}
    \subfloat[\label{subfig:qnli2}]{
         \includegraphics[width=0.482\linewidth]{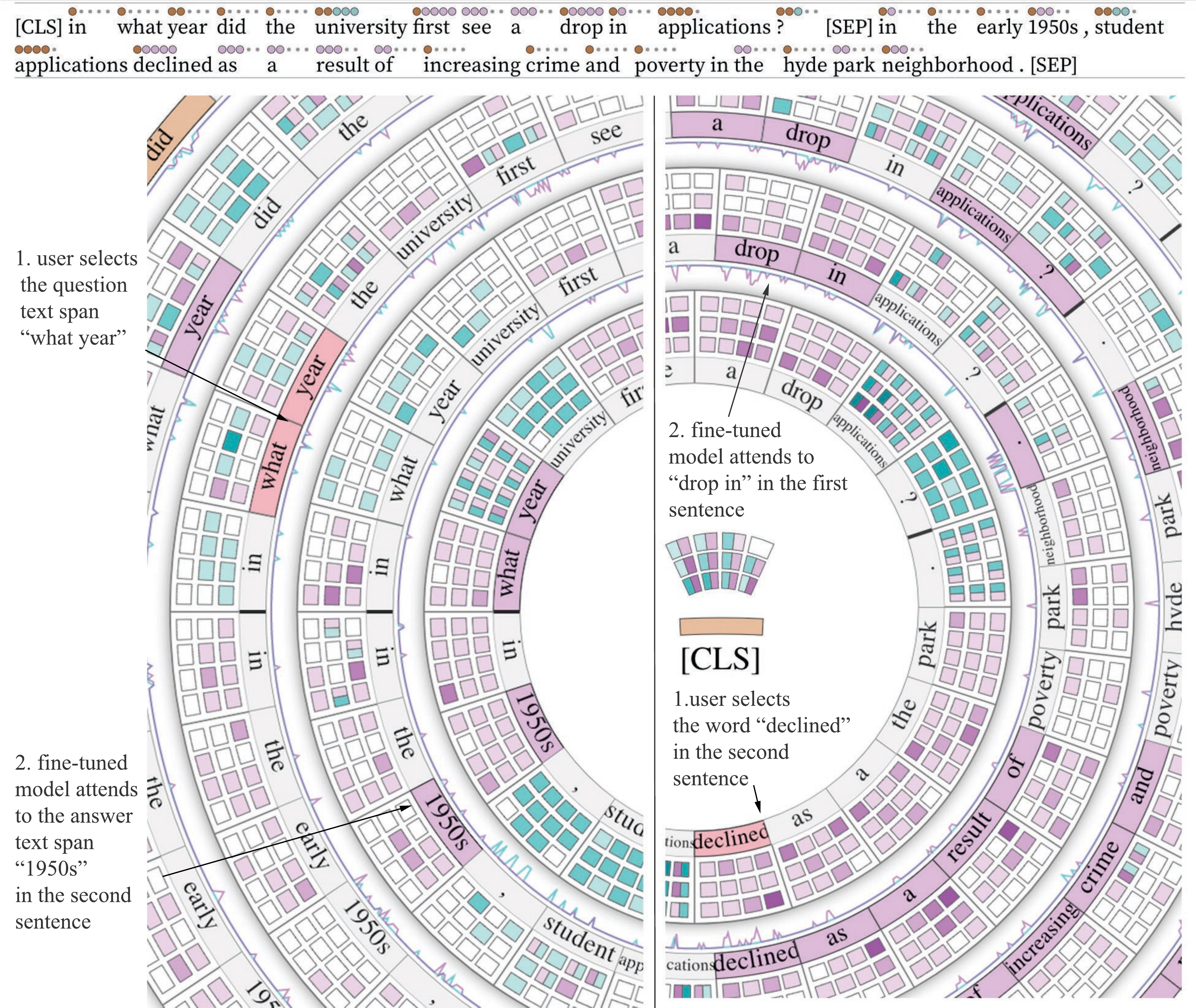}
	}
    \caption{Here we show examples for recognizing question answer pairings. In (a), the answer span (``more than 50\%'') attends to ``what'', followed by the full question span. Conversely (b), selecting ``what year'' shows a dependency for the answer span (``1950s'') in the proceeding layer.}
    \label{fig:qnli}
\end{figure*}

\begin{figure*}[!t]
    \subfloat[\label{subfig:mrpc1}]{
         \includegraphics[width=0.5\linewidth]{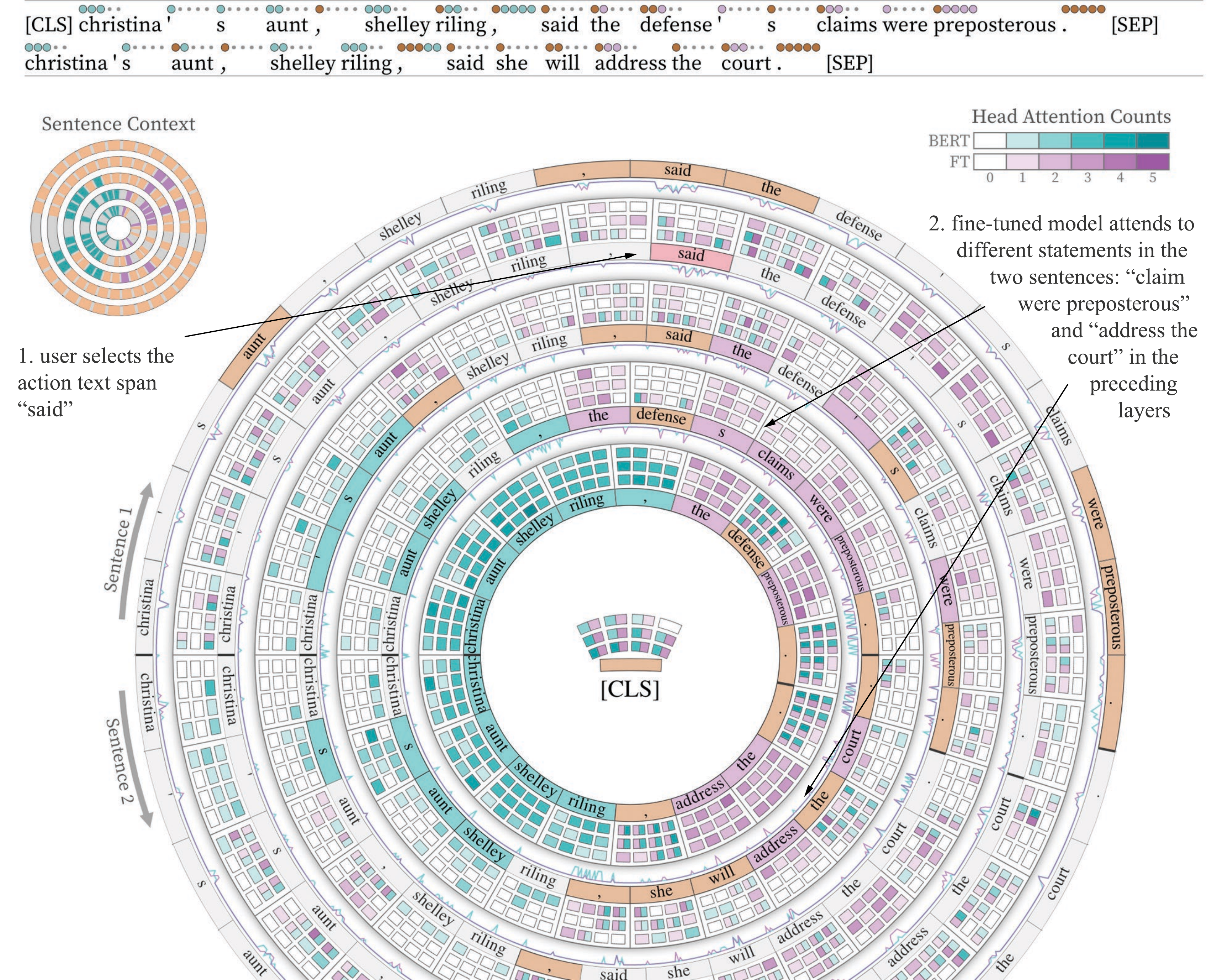}
	}
    \subfloat[\label{subfig:mrpc2}]{
         \includegraphics[width=0.5\linewidth]{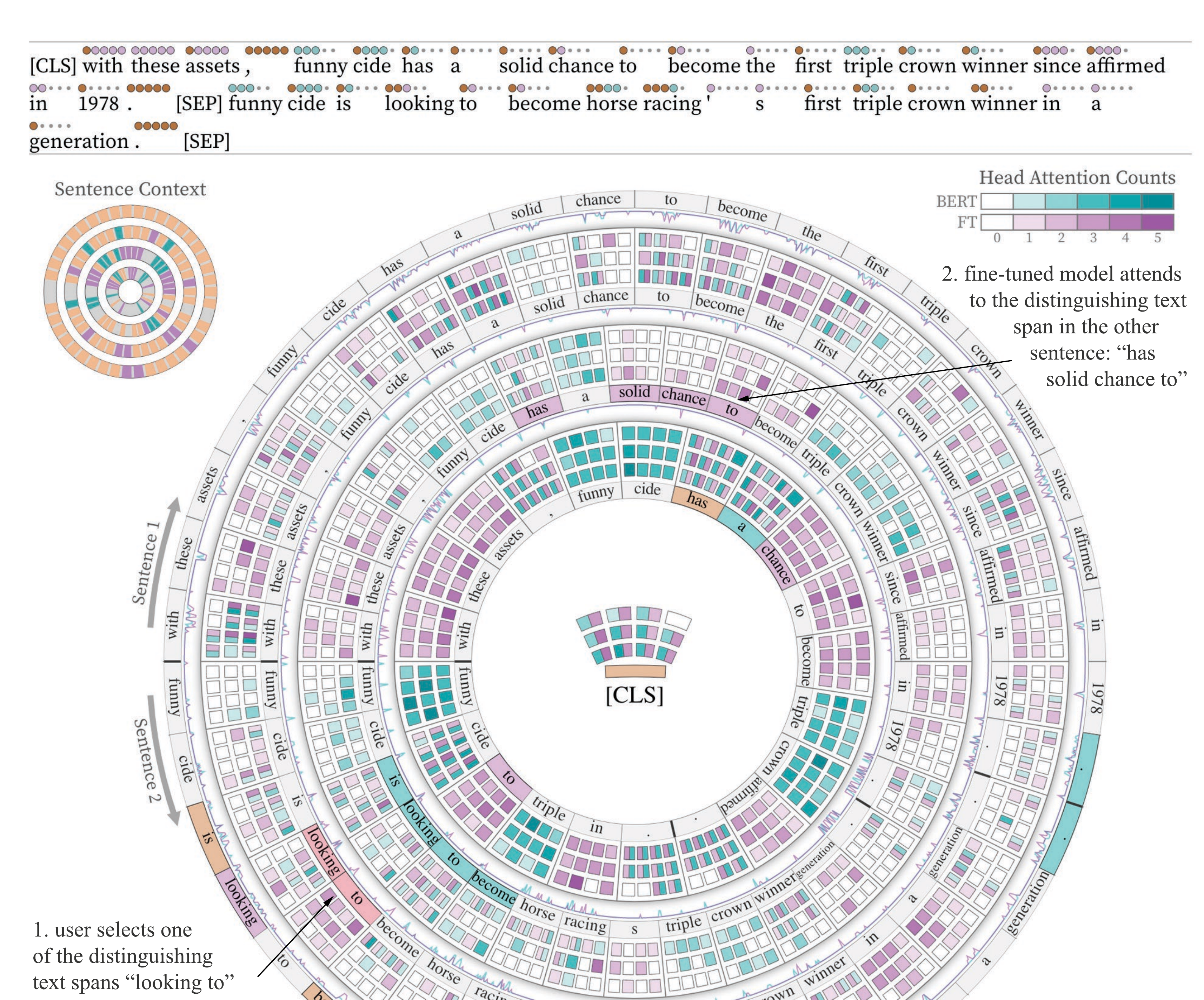}
	}
    \caption{We demonstrate our interface for recognizing that one sentence fails to paraphrase another. In (a), we find that selecting ``said'' has a dependency on the two distinguishing statements: ``claims were preposterous'' and ``address the court''. In (b), selecting one of the distinguishing text spans ``looking to'' shows attention flowing to the other distinguishing span: ``solid chance to''.}
    \label{fig:mrpc}
\end{figure*}

\textbf{RTE} (Recognizing Textual Entailment) asks the model if the second sentence is an entailment of the first. For example, the sentence-pair \textit{[CLS]Mount Olympus towers up from the center of the earth. [SEP] Mount Olympus is in the center of the earth. [SEP]} is classified positive since the second sentence can be inferred from the first sentence.

\textbf{QNLI} (Question Natural Language Inference) asks the model to determine if the second sentence contains the correct answer to the question in the first sentence. For example, the sample \textit{[CLS] What percentage of this farmland grows wheat? [SEP] More than 50\% of this area grows wheat. [SEP]} is classified as positive.

\textbf{MRPC} (Microsoft Research Paraphrase Corpus) asks the model to determine if two sentences have the same meaning. For example, the pair \textit{[CLS] It affected earnings per share by a penny. [SEP] The company said this impacted earnings by a penny a share. [SEP]} is classified positive since the second sentence paraphrase the first.

\subsection{Use Case: Textual Entailment}

In Fig.~\ref{fig:rte_use_case} we show an example of how to use Attention Flows to understand differences learned during fine-tuning for textual entailment. In this example, the first sentence entails the hypothesis found in the second, as ``will receive the proceeds'' implies ``proceeds go to'', both in reference to family. Hence, the user selects the token ``go'' at the penultimate layer and ``receive'' at an earlier layer, to understand the model's attention flows between these tokens. The proceeding layer from the ``receive'' selection highlights how both models attend to tokens that belong to the first sentence. However, at the next layer, we see that the fine-tuned model's attention crosses to the second sentence, picking up on the relevant text span to determine entailment. This layer, and the next, gathers the necessary tokens for the task (``family'' then ``hepburn''), before arriving at ``go''. Note that BERT's pre-trained attention contains heads for the aforementioned layer's distinguishing text span, but for this specific selection, only the heads that correspond to the fine-tuned model are enacted. Furthermore, we observe that the \emph{same} head in the fine-tuned model is used for this text span.

In Fig.~\ref{fig:rte2} we show an example where the hypothesis in the second sentence does not lead to entailment. The user selects ``doha'' from the entailed sentence, and selects ``qatar'' from the entailed (a) and entailing (b) sentences to understand the relationship between these entities via attention flows from ``qatar'' to ``doha''. Here, we find that the fine-tuned model places more importance on the past participle ``located in'', compared with the pre-trained model, and thus ``doha'' is likely to be more informed by this description. Further, we find that ``capital city'' is attended to in the fine-tuned model, but not the pre-trained model, and this is a key phrase that determines the absence of entailment.

From these examples, and others, we make several observations. First, we find that the pre-trained model tends to focus on nouns (``hepburn'', ``family'', ``sale''), while the fine-tuned model only does so if it is relevant to the task (``qatar'' and ``doha''). Further, we find that the fine-tuned model tends to focus on prepositions (``to'', ``from'', ``of''), verbs (``receive'', ``located in''), and possessive phrases. These parts of speech help establish relationships between nouns, e.g. named entities, and are common to the hypotheses involved in entailment, e.g. whether or not two entities are performing the same task.

\subsection{Use Case: Question Answer Verification}

Fig.~\ref{fig:qnli} shows an example of model comparison for the task of question answer verification. In Fig.~\ref{fig:qnli}(a), the user first selects the answer ``more than 50\%''. Inspecting its dependent tokens, we see that the immediately preceding layer corresponds to the question of ``what'', and the layer further back highlights the remainder of the text span of the question. Hence, we see that the answer text span is able to gather the appropriate question information through multi-layer reasoning. Fig.~\ref{fig:qnli}(b)-left considers the converse: if we select the tokens that identify the question (``what year''), will this be made available to the answer span? Indeed, we find that the object of the answer in the next layer, ``1950s'', depends on ``what year''. On the other hand, the pre-trained model fails to attend to the answer across multiple layers, as indicated by most attention heads colored purple for the answer span. Further, on the right side of Fig.~\ref{fig:qnli}(b), selecting the token ``declined'' in the answer shows that it depends on ``drop in applications'' from the question in preceding layers, and these phrases link the two sentences together in reference to ``applications''.

These two examples highlight a pattern in many cases of QNLI where the classification is positive (e.g. the second sentence \emph{does} answer the question in the first sentence), namely that the question aspect of the first sentence (e.g. one of the ``5 W's'') serves as a bridge to the second sentence and frequently, the text span corresponding to the answer. We note that the answer spans are not technically required for answering the question, yet this information is nevertheless learned via fine-tuning, indicating that the model is not merely picking up on task-irrelevant details to solve the problem.

\subsection{Use Case: Paraphrasing Verification}

Last, we analyze fine-tuning for the problem of paraphrase verification, please see Fig.~\ref{fig:mrpc}. For the negative (e.g. not a paraphrasing) example in Fig.~\ref{fig:mrpc}(a), the user first selects the token ``said'', and we can observe that the fine-tuned model attends to what the person said in \emph{both} sentences. Moreover, only the fine-tuned model attends to the tokens that differentiate the sentences, e.g. ``claim were preposterous'' and ``address court''. Fig.~\ref{fig:mrpc}(b) highlights another negative paraphrasing example, where the user brushes the text span ``looking to'' in the second sentence. We find that the fine-tuned model attends to the text span in the first sentence from the proceeding layer -- ``solid chance to'' -- that distinguishes the sentences. Note, the pre-trained model does not attend to any tokens in the first sentence, instead expanding attention to nearby tokens. We also find that the [CLS] token for the pre-trained model tends to place importance on matched text spans between the sentences, e.g. ``funny cide'', ``triple crown'', whereas the fine-tuned model seeks relevant, distinguishing tokens. For instance, ``horse racing'' is not found in the first sentence, but this phrase does not distinguish the two sentences, and in fact, the pre-trained model places more importance on this phrase as shown in the sentence view.

For the MRPC task, our experiments show that, for negative samples, the fine-tuned BERT model attends to words that distinguish the two sentences while the pre-trained BERT model does not. Existing research~\cite{Brunner2020On} claims that, because the BERT model computes word embeddings that are contextualized with regards to the sentence, the model will learn increasingly abstract representations as the sentence goes through deeper layers, and thus, word embeddings in the last few layers will lose their identity. However, our experiments show that, in the case of the sentence-pair paraphrasing task, the BERT model still retains the identities of the tokens in deeper layers, which enables the model to attend to words of the two sentences that distinguish them.

\subsection{User Feedback}

To verify the effectiveness of our design, we collected user feedback from two different groups. The first group consisted of experts in NLP, specifically three researchers within academia, where we aimed to answer the following: for domain experts, is our interface intuitive, easy to use, and helpful for model comparison? The second group consisted of a broader population, where we crowdsourced participants (total of seven) with a self-reported undergraduate education in Computer Science, aiming to answer the following: for potential non-experts, how effective is our interface for assessing differences between models?

To gather feedback, we directly recruited NLP experts, while we used the crowdsourcing platform Prolific\footnote{\url{https://prolific.co}} to recruit the broader audience. In both cases, we provided a brief explanation of the interface, provided a set of sentences from the above 3 tasks, and asked participants to freely use the interface. Participants were then asked to provided survey responses on: (a) their findings in using the interface, and (b) the usability of the interface. We summarize the group-specific findings:

\textbf{NLP Experts:} Overall, the NLP researchers enjoyed the visualization. One researcher mentioned that they understood the visual encodings ``without having to read the instructions'', and found the interface useful for identifying differences between pre-training and fine-tuning. Another researcher found the sentence summary useful for providing overviews but acknowledged that the central visualization was rather complex to understand. The third researcher liked how the interface compares ``pre-trained and fine-tuned models via colors, it helps show patterns quickly'', and expressed interest in using the interface for more general model comparisons, beyond fine-tuning.

\textbf{Crowdsourced Participants:} In using the interface, the participants all found the fine-tuned model attended to task-relevant details better than the pre-trained model, specifically, dealing with ``medium-sized sentences'', better handling second sentences in tasks ``in terms of head attention counts'', recognizing ``importance of numbers'', handling ``verbs and objects in sentences'', and handling actions where ``someone was doing something''. One participant thought, in contrast, that the pre-trained model was ``gibberish-ish all the way''. Regarding usability, two participants mentioned that it took a while to understand, but eventually, became easy to use. One participant mentioned that toggling views made it easier to comprehend the main view, and ``highlighting tokens was somewhat helpful in understanding the token selection process.'' Other participants, however, did find certain aspects difficult to use, with the purpose of certain interactions not being evident.

%% file: futurework.tex
We introduced Attention Flows: a tool that supports exploration into how self-attention of Transformer models is refined during fine-tuning, and how self-attention informs classification decisions. Through use cases of our interface, we show  how self-attention evolves to address task-specific details, while our user feedback validates the usability, and potential insights, provided by the interface.

We plan to explore several avenues of future work. In our design, we only consider self-attention matrices and do not consider other aspects of the Transformer model, such as contextualized embeddings. We plan on linking attention functionality with word embeddings to provide a more comprehensive view. Further, we plan to extend our approach to more general model comparison, e.g. models of different pre-training objectives~\cite{joshi2020spanbert} or optimization schemes~\cite{liu2019roberta}.

Our tool supports the comparison of a single sample across models, but a limitation with our current interface is how such samples are selected. We offer a simple interface for browsing samples according to the model's classification, yet a way to summarize and browse a collection of samples would be more useful for the end user. For future work, we intend to use our scheme for identifying attention heads as a means of querying sentence pairs that contain similar patterns in self-attention, in order to find and compare different samples. Such an interface can facilitate the investigation of existing benchmark datasets~\cite{wang2018glue}, ensuring that a user's discoveries holds across many samples, as a means to identify shortcomings in pre-training or fine-tuning.